\documentclass[useAMS,usenatbib]{mn2e}

\usepackage{graphicx}
\usepackage{url}
\usepackage{bm}
\usepackage{amsmath}
\usepackage{amssymb}

\title[Impossibility of testing ISL by TTVs]{On the (im)possibility of testing new physics in exoplanets using transit timing variations: deviation from inverse-square law of gravity}
\author[Y. Xie \& X.-M. Deng]{Yi Xie$^{1,3}$\thanks{E-mail:yixie@nju.edu.cn} and Xue-Mei Deng$^{2,3}$\\
$^{1}$School of Astronomy and Space Science, Nanjing University, Nanjing 210093, China\\
$^{2}$Purple Mountain Observatory, Chinese Academy of Sciences, Nanjing 210008, China\\
$^{3}$Key Laboratory of Modern Astronomy and Astrophysics, Nanjing University, Ministry of Education, Nanjing 210093, China}
\begin{document}

\date{Accepted . Received ; in original form }

\pagerange{\pageref{firstpage}--\pageref{lastpage}} \pubyear{2002}

\maketitle

\label{firstpage}

\begin{abstract}
Ground-based and space-borne observatories studying exoplanetary transits now and in the future will considerably increase the number of known exoplanets and the precision of the measured times of transit minima. Variations in the transit times can not only be used to infer the presence of additional planets, but might also provide opportunities for testing new physics in the places beyond the Solar system. In this work, we take deviation from the inverse-square law of gravity as an example, focus on the fifth-force-like Yukawa-type correction to the Newtonian gravitational force which parameterizes this deviation, investigate its effects on the secular transit timing variations and analyze their observability in exoplanetary systems. It is found that the most optimistic values of Yukawa-type secular transit timing variations are at the level of $\sim 0.1$ seconds per year. Those values unfortunately appear only in rarely unique cases and, most importantly, they are still at least two orders of magnitude below the current capabilities of observations. Such a deviation from the inverse-square law of gravity is likely too small to detect for the foreseeable future. Meanwhile, systematic uncertainties, such as the presence of additional and unknown planets, will likely be exceptionally difficult to remove from a signal that should be seen.

\end{abstract}


\section{Introduction}
\label{sec:intro}

Currently, more than 880 exoplanets have been discovered and about 300 of them are in the transiting systems.\footnote{\url{http://exoplanet.eu/catalog/}} Now and in the future, ground-based and space-borne observatories used for studying transits of exoplanets will considerably increase the number of known exoplanets and the precision of the observed times of transit minima.\footnote{As pointed out by \cite{Kipping2011PhDT}, the so-called ``mid-transit time'' in the exoplanet literature is highly ambiguous. Following the terminology in \cite{Kipping2011PhDT}, we will use ``transit minimum'' and ``time of transit minimum'' in this paper. Because, for a limb-darkened star, the transit minimum occurs when the apparent sky-projected separation between the exoplanet and the star reaches a minimum, which has a completely unambiguous definition. So ``transit timing variations'' also refers to ``changes of times of transit minimum''.} The measured transit timing variations (TTVs) can be used to infer the presence of additional planets \citep[e.g.][]{Holman2005Sci307.1288,Agol2005MNRAS359.567,Heyl2007MNRAS377.1511,Nesvorny2012Sci336.1133} and study the dynamics of multiple planets systems \citep[e.g.][]{Holman2010Sci330.51,Lissauer2011Nature470.53,Fabrycky2012ApJ750.114,Ford2012ApJ750.113,Steffen2012MNRAS421.2342,Nesvorny2013arXiv1304.4283}. Recently, the \textit{Kepler} mission \citep{Koch2010ApJ713.L79,Borucki2010Sci327.977} released a catalog of transit timing measurements of the first twelve quarters, which identifies the Kepler objects of interest with significant TTVs \citep{Mazeh2013ApJS208.16}. Such a large amount of confirmed and potential transiting exoplanets provide opportunities for testing physical laws of nature, especially fundamental theories of gravity, in the places beyond the Solar system.

But, what is the necessity of performing these tests in exoplanetary systems? After all, modified and alternative relativistic theories of gravity have been tested in the Solar system with very high precision \citep[see][for reviews]{Will1993TEGP,Will2006LRR9.3,Turyshev2008ARNPS58.207}, whereas tests in exoplanetary systems in current stage are expected to be much worse for lack of high-accuracy observations. However, some observations indicate the fundamental constants of nature might have a temporal and spatial variation \citep[e.g.][]{Webb1999PRL82.884,Murphy2001MNRAS327.1223,Murphy2001MNRAS327.1237,Murphy2001MNRAS327.1244,Murphy2001MNRAS327.1208, Webb2001PRL87.091301,Murphy2003MNRAS345.609,Murphy2007MNRAS378.221,Murphy2008MNRAS384.1053,Webb2011PRL107.191101}, which imply some fundamental laws of nature may also have variations in such a manner although it ought be very small. Furthermore, scalar-tensor theories of gravitation, as alternatives of general relativity (GR), also theoretically imply these subtle variations may exist even in the scale of planetary systems due to the possible couplings between matter and scalar fields \cite[see][for a review]{Fujii2007BookSTT}. In order to (dis)prove it empirically, we need to go to different times and places. Exoplanetary systems can serve as testbeds outside the Solar system for conducting those tests owning to their unique locations. Therefore, in this work, we will focus on testing the inverse-square law (ISL) of gravity.

A way to parameterize deviation from the ISL of gravity, which might be caused by new physics beyond the standard model of particles and GR, is the fifth-force-like Yukawa-type correction to the Newtonian gravitational force, which has been intensively studied \citep[e.g.][]{Fischbach1986,Fischbach1992,Iorio2002PhLA298.315,Adelberger2003,Iorio2007JHEP10.041,Lucchesi2010PRL105.231103,Haranas2011ApSS331.115,Haranas2011ApSS332.107,Lucchesi2011AdSR47.1232,Iorio2012JHEP05.073,Deng2013MNRAS431.3236}. The gravitational potential with this correction is
\begin{equation}
  \label{}
  V = V_{\mathrm{N}}(r)+V_{\mathrm{YK}}(r),
\end{equation}
where the Newtonian potential and Yukawa-type correction are, respectively,
\begin{eqnarray}
  \label{}
  V_{\mathrm{N}}(r) & = & \frac{GM_1M_2}{r},\\
  \label{VYK}
  V_{\mathrm{YK}}(r) & = & \frac{GM_1M_2}{r}\alpha \exp\bigg(-\frac{r}{\lambda}\bigg).
\end{eqnarray}
Here $G$ is the gravitational constant, $M_i$ $(i=1,2)$ is the mass of the $i$th body and $r$ is the distance between them. $\alpha$ is a dimensionless strength parameter of the correction and $\lambda$ is a length scale for it \citep[see][for a review of constraints on $\alpha$ and $\lambda$]{Fischbach1999}. Under this parameterization, if the inverse-square law of gravity is violated, an extra force will exert on exoplanets and cause \emph{additional} secular TTVs.

Among mainly known sources of secular TTVs, the general relativistic periastron advance (GRPA) contributes. Its observability in exoplanets has been investigated in several works \citep[e.g.][]{Miralda-Escude2002ApJ564.1019,Adams2006ApJ649.992,Adams2006ApJ649.1004,Adams2006IJMPD15.2133,Iorio2006NewA11.490,Heyl2007MNRAS377.1511,Jordan2008ApJ685.543,Pal2008MNRAS389.191,Li2010ApSS327.59,Iorio2011MNRAS411.167,Iorio2011ApSS331.485,Li2012ApSS341.323,Zhao2013RAA13.1231}. It is found that GRPA can be detectable on timescales of less than about 10 years with current observational capabilities by observing the times of transits in exoplanets \citep{Jordan2008ApJ685.543}.

This means that, like the well-known phenomena in the Solar system, such as the anomaly in the perihelion shift of Mercury \citep{Nobili1986Nature320.39} that gave a hint at new physics about GR and the dynamics of planets which could be used to test fundamental laws of physics \citep[e.g.][]{Iorio2005A&A431.385,Iorio2005A&A433.385,Iorio2005JCAP07.008,Iorio2005JCAP09.006,Iorio2006JCAP08.007,Ruggiero2007JCAP01.010,Folkner2010IAUS261.155,Pitjeva2010IAUS261.170,Fienga2011CMDA111.363,Iorio2010JCAP06.004,Iorio2011ApSS331.351,Iorio2011JCAP05.019,Iorio2012SoPh281.815,Iorio2012JCAP07.001,Pitjeva2012IAUJD7.37,Pitjev2013AstL39.141,Pitjeva2013MNRAS432.3431,Xie2013MNRAS433.3584}, observing of secular TTVs can also serve as a testbed with the help of high-precision measurements which might be available in the future. It will also provide opportunities to test the fundamental theories of gravity in quite a large number of different and unique locations beyond the Solar system. This will make transiting exoplanets very similar to binary pulsars in testing physical laws describing gravity \citep[e.g.][]{Bell1996ApJ464.857,Damour1996PRD53.5541,Kramer2006Sci314.97,Iorio2007ApSS312.331,Deng2009PRD79.044014,Li2010ApSS327.59,Deng2011SCG54.2071,Li2011ApSS334.125,Laurentis2012MNRAS424.2371,Ragos2013ApSS345.67,Xie2013RAA13.1}.

Hence, we will investigate the (im)possibility of detecting fifth-force-like Yukawa-type effects on the secular TTVs as an example of trying to test new physics in exoplanets. After analysing their observability in exoplanetary systems, we find that the most optimistic values of this type secular TTVs are at the level of $\sim 0.1$ seconds per year. Those values unfortunately appear only in rarely unique cases and, most importantly, they are still at least two orders of magnitude below the current capabilities of observations. Such deviation from the ISL of gravity is likely too small to detect for the foreseeable future. Meanwhile, systematic uncertainties, such as the presence of additional and unknown planets, will likely be exceptionally difficult to remove from a signal that should be seen. 

The rest of the paper is organized as follows. Section \ref{sec:YukawaTTV} is devoted to describing TTVs under the Yukawa-type correction. In section \ref{sec:obs}, we present an analysis about its observability in the secular TTVs. Finally, in section \ref{sec:con}, we summarize our results.

\section{TTVs caused by Yukawa-type correction}

\label{sec:YukawaTTV}

To describe the dynamics of a transiting exoplanetary system, understand its transit light curve and represent the observables, we adopt the coordinate systems defined and applied in \cite{Kipping2011PhDT}. The plane of $\hat{X}$-$\hat{Y}$ is defined as the plane of the sky, the host star is at the origin $O$ and the observer is located at $(X,Y,Z)=(0,0,+\infty)$. Then , in the $\hat{X}$-$\hat{Y}$-$\hat{Z}$ system, the inclination of a transiting exoplanet $i$ is close to $90^{\circ}$. Hereafter, we take widely used notations in celestial mechanics: $a$ is the semi-major axis, $e$ is the eccentricity, $i$ is the inclination, $\Omega$ is the longitude of the ascending node, $\omega$ is the argument of periastron, $M$ is the mean anomaly and $f$ is the true anomaly. The normalized apparent (sky-projected) separation between the planet and the star is defined as \citep{Kipping2011PhDT}
\begin{eqnarray}
  \label{}
  S & \equiv & \frac{1}{R_{\ast}}\sqrt{X^2+Y^2} \nonumber\\
  & = & \frac{a}{R_{\ast}}\varrho(f)\sqrt{1-\sin^2(\omega+f)\sin^2i},
\end{eqnarray}
where $\varrho(f) \equiv (1-e^2)/(1+e\cos f)$ and $R_{\ast}$ is the radius of the star. For mathematical convenience in the following parts of this paper, we will also use the expression for $S^2$:
\begin{equation}
  \label{}
  S^2 = \frac{a^2}{R_{\ast}^2}\varrho^2(f)[1-\sin^2(\omega+f)\sin^2i].
\end{equation}

\subsection{Transit minima}

For a Kelperian transiting exoplanet, the instants of transit minima (and maxima) occur when $\mathrm{d}S/\mathrm{d}t=0$ \citep{Kipping2011PhDT}, which leads to
\begin{equation}
  \label{dSdt=0}
  \frac{\mathrm{d}S}{\mathrm{d}t} = \frac{\mathrm{d}S}{\mathrm{d}f} \frac{\mathrm{d}f}{\mathrm{d}t} = 0.
\end{equation}
It is worth mentioning that condition defined by equation (\ref{dSdt=0}) is a pure geometric criterion without any ambiguity. Because $\mathrm{d}f/\mathrm{d}t \neq 0$ for planetary orbital motions, the condition $\mathrm{d}S/\mathrm{d}t=0$ is equivalent to $\mathrm{d}S/\mathrm{d}f=0$. An easier way to handle the mathematics is to make use of $\mathrm{d}S^2/\mathrm{d}f=0$, and such a condition can be proven to be equivalent to $\mathrm{d}S/\mathrm{d}f=0$ \citep{Kipping2011PhDT}. To obtain the true anomaly at the transit minima $f_T$ (the subscript ``T'' denotes transit), one needs to solve a quartic equation involving $\cos f$ [see eq. (4.5) in \citet{Kipping2011PhDT}]. Although solving it is mathematically possible, the solutions are pretty lengthy and impractical. Treating $\cos^2i$ as a small quantity which is very close to zero for transiting exoplanets and using the Newton-Raphson iteration method, Kipping shows the series expansion solution for $f_T$ can be written as \citep{Kipping2011PhDT}
\begin{equation}
  \label{fT}
  f_{T} = \bigg[ \frac{\pi}{2} - \omega \bigg] -\sum_{j=1}^{n} \eta^T_j,
\end{equation}
where, by $h \equiv e\sin \omega$ and $k \equiv e\cos \omega$,
\begin{eqnarray}
  \label{}
  \eta^T_1 & = & \bigg(\frac{k}{1+h}\bigg)(\cos^2i)^1,\\
  \eta^T_2 & = & \bigg(\frac{k}{1+h}\bigg)\bigg(\frac{1}{1+h}\bigg)(\cos^2i)^2,\\
  \eta^T_3 & = & \bigg(\frac{k}{1+h}\bigg)\bigg[\frac{6(1+h)+k^2(1-2h)}{6(1+h)^3}\bigg](\cos^2i)^3.
\end{eqnarray}
It is demonstrated \citep{Kipping2011PhDT} that using a first-order expansion solution can reduce the error to less than a millisecond for a highly eccentric planet with a short period. Solutions expanded to higher orders (up to $n=6$) can be found in \citet{Kipping2011PhDT}.

\subsection{Yukawa-type secular TTVs}

With the same approach used in \cite{Iorio2011MNRAS411.167} to work out long-term time variations of some observables for transiting exoplanets, for a given observable $\Gamma$ that is a function of Keplerian orbital elements, i.e. $\Gamma=\Gamma(\{\sigma\})$, where $\{\sigma\}=\{a,e,i,\Omega,\omega,M\}$, if perturbations on the Keperlian orbital motion are taken into account, we can calculate its secular variation by averaging:
\begin{equation}
  \label{}
  \bigg<\frac{\mathrm{d}\Gamma}{\mathrm{d}t}\bigg> =  \frac{1}{P}\int_0^P \frac{\mathrm{d}\Gamma}{\mathrm{d}t} \mathrm{d}t  = \frac{1}{P}\int_0^P \sum_{\kappa\in\{\sigma\}}\frac{\partial\Gamma}{\partial \kappa}\frac{\mathrm{d}\kappa}{\mathrm{d}t} \mathrm{d}t,
\end{equation}
where $P$ is the Keplerian period of the orbit. Applying this approach to $f_T$, we can obtain its secular changes as
\begin{equation}
  \label{<dftdt>}
  \bigg<\frac{\mathrm{d}f_T}{\mathrm{d}t}\bigg> = -\,\bigg<\frac{\mathrm{d}\omega}{\mathrm{d}t}\bigg> -\sum_{j=1}^{n} \bigg<\frac{\mathrm{d}\eta^T_j}{\mathrm{d}t}\bigg>.
\end{equation}
If a two-body problem with the fifth-force-like Yukawa-type correction is considered, we can find that
\begin{eqnarray}
  \label{}
  \bigg<\frac{\mathrm{d}\eta^T_1}{\mathrm{d}t}\bigg> & = &  -\frac{h+e^2}{(1+h)^2} \cos^2i \bigg<\frac{\mathrm{d}\omega}{\mathrm{d}t}\bigg>,\\
  \bigg<\frac{\mathrm{d}\eta^T_{2}}{\mathrm{d}t}\bigg> & = &  -\frac{k^2+h+e^2}{(1+h)^3} \cos^4i \bigg<\frac{\mathrm{d}\omega}{\mathrm{d}t}\bigg>,\\
  \bigg<\frac{\mathrm{d}\eta^T_{3}}{\mathrm{d}t}\bigg> & = &   \frac{1}{2} (1+h)^{-5}(2 h^3 k^2+2 h k^4+h^2 k^2-2 k^4-2 h^3\nonumber\\
  & & -7 h k^2-4 h^2-6 k^2-2 h) \cos^6i \, \bigg<\frac{\mathrm{d}\omega}{\mathrm{d}t}\bigg>. 
\end{eqnarray}
Here, the secular variations of $a$, $e$, $i$ and $\Omega$ are zero  and only $\left<\mathrm{d}\omega/\mathrm{d}t\right>$ contributes in the above ones [see eqs. (18)--(22) in \cite{Deng2013MNRAS431.3236}] because
\begin{equation}
  \label{Yukawadomegadt}
  \bigg<\frac{\mathrm{d} \omega}{\mathrm{d} t}\bigg>  =  \alpha \frac{n a\sqrt{1-e^2}}{e\lambda}\exp\bigg(-\frac{a}{\lambda}\bigg) I_1\bigg(\frac{ae}{\lambda}\bigg),
\end{equation}
where $I_1(z) = \mathrm{d}I_0(z)/\mathrm{d}z$ and $I_0(z)$ is the modified Bessel function of the first kind \citep{Arfken2005}.

However, the secular variation of $f_T$ is not an observable practically so that, for realistic measurements, it needs to be converted to the secular variation of time of transit minimum $t_T$, i.e. secular TTV,
\begin{equation}
  \label{}
   \bigg<\frac{\mathrm{d}t_T}{\mathrm{d}t}\bigg>  =   \bigg<\frac{\mathrm{d}t_T}{\mathrm{d}f_T}\frac{\mathrm{d}f_T}{\mathrm{d}t}\bigg> = \frac{1}{n\sqrt{1-e^2}}{\varrho}_T^2  \bigg<\frac{\mathrm{d}f_T}{\mathrm{d}t}\bigg>,
\end{equation}
where $\varrho_T \equiv \varrho(f_T)$ and $n=2\pi/P$. By substituting equations (\ref{fT}), (\ref{<dftdt>}) and (\ref{Yukawadomegadt}) into above one, we can have
\begin{eqnarray}
  \label{}
  \bigg<\frac{\mathrm{d}t_T}{\mathrm{d}t}\bigg> & = & -\frac{1}{n\sqrt{1-e^2}}{\varrho}_T^2  \, \bigg<\frac{\mathrm{d}\omega}{\mathrm{d}t}\bigg> + \mathcal{O}(\cos^2i)\nonumber\\
  & = & -\frac{\alpha}{e} \frac{a}{\lambda} {\varrho}_T^2 \exp\bigg(-\frac{a}{\lambda}\bigg) I_1\bigg(\frac{ae}{\lambda}\bigg) + \mathcal{O}(\cos^2i).
\end{eqnarray}
For a time duration $\Delta t$, the Yukawa-type secular TTV $\Delta t_T$ is
\begin{equation}
  \label{DtTDt}
  \frac{\Delta t_T}{\Delta t} = -\frac{\alpha}{e} \xi {\varrho}_T^2 \exp(-\xi) I_1(\xi e) + \mathcal{O}(\cos^2i),
\end{equation}
where
\begin{equation}
  \label{defvarrhoT}
  \varrho_T=(1-e^2)/(1+e\sin\omega) + \mathcal{O}(\cos^2i),
\end{equation}
and
\begin{equation}
  \label{defxi}
  \xi\equiv a/\lambda.
\end{equation}
It is worth mentioning that although equation (\ref{DtTDt}) does not diverge when $e=0$ because
\begin{equation}
  \label{}
  \lim_{e\rightarrow0} \frac{1}{e}I_1(\xi e) = \frac{\xi}{2},
\end{equation}
it is not suitable for the case that $e$ is extremely close to $0$ because it makes $\omega$ ill-defined. In order to avoid this, one needs to reformulate it in terms of singularity-free orbital elements \citep{Danby1962Book}. While the orbit may precess rapidly near zero eccentricity, its effect would simply be a change to the measured orbital period (yielding an orbital frequency of $2 \pi/ P_{\mathrm{measured}} = 2 \pi / P_{\mathrm{true}} + \dot{\omega}$). The TTV signal in this case is the deviation from a constant interval between transits. Thus, the changing angular velocity of a planet on an elliptical orbit is an essential aspect of the signal. However, due to the requirement of an elliptical orbit, short-period planets (with periods of only a few days) are not likely to be good candidates because of the circularization of the orbit from tides. It can also be found that, from equation (\ref{defvarrhoT}),
\begin{equation}
  \label{estvarrho}
  (1-e)^2 + \mathcal{O}(\cos^2i) \le \varrho_T^2 \le (1+e)^2 + \mathcal{O}(\cos^2i),
\end{equation}
which makes ${\Delta t_T}/{\Delta t}$ not sensitive to $\omega$ according to equation (\ref{DtTDt}). And, since $\xi$ is a dimensionless parameter as a ratio of the semi-major axis $a$ and the length scale of Yukawa correction $\lambda$, it suggests that it will be very difficult to determine $\lambda$ by observations on TTVs alone. In ${\Delta t_T}/{\Delta t}$, $\alpha$ and $e$ play more important roles (see next section for details).

Equation (\ref{DtTDt}) describes changes of times of transit minimum for a time duration which may be much longer than the Keplerian period of the orbit. In practice, an important directly measurable quantity is the temporal interval $P_T$ between successive transits. It will change as well because of $\dot{\omega}$. This variation gives the observed deviations from a linear ephemeris. Up to the first order of $e$, the derivative of the transit period $P_T$ is given by \citep{Miralda-Escude2002ApJ564.1019,Heyl2007MNRAS377.1511,Jordan2008ApJ685.543}
\begin{equation}
  \label{}
  \dot{P}_T  =  4\pi e \bigg(\frac{\dot{\omega}}{n}\bigg)^2 \sin(M_T)
\end{equation}
so that the contribution caused by Yukawa-type correction is
\begin{equation}
  \dot{P}_T  =  4\pi \alpha^2 \xi^2 e^{-1}(1-e^2)\exp(-2\xi)I^2_1(\xi e)\sin(M_T),
\end{equation}
where $M_T$ is the mean anomaly at transit \citep{Miralda-Escude2002ApJ564.1019} and is related to the true anomaly at transit $f_T$ of the first order of $e$ by $M_T=f_T-2e\sin f_T$.

\section{Observability of Yukawa-type secular TTVs}

\label{sec:obs}

This section will be dedicated to an important issue: the observability of these Yukawa-type effects. Based on the literature about Yukawa-type correction \citep[e.g.][]{Fischbach1999,Adelberger2003,Deng2009PRD79.044014,Lucchesi2010PRL105.231103,Deng2011SCG54.2071,Lucchesi2011AdSR47.1232,Iorio2012JHEP05.073} and the catalog of confirmed transiting exoplanets \footnote{\url{http://exoplanet.eu/catalog/}}, we will focus on the domain of the parameters in equation (\ref{DtTDt}) as
\begin{eqnarray}
  \label{}
  \mathcal{D} & = & \{(\xi,\alpha,e,\omega) |\, 10^{-2}\le\xi\le10^1,\,10^{-12}\le\alpha\le10^{-8},\nonumber\\
  & &\phantom{\{(\xi,\alpha,e,\omega) |\,} 0.01\le e \le 0.6,\,0\le \omega\le2\pi \}.
\end{eqnarray}
In the construction of this space of parameters, we take confirmed transiting exoplanets as samples of orbital configurations. We also consider constraints on Yukawa correction in the Solar system. For instance, one of the upper bounds of Yukawa correction is given by \cite{Iorio2007JHEP10.041} using the Solar system planets orbital motion with EPM2004 ephemeris \citep{Pitjeva2005SoSyR39.176}: $\alpha \sim 10^{-9}$ and $\lambda \sim 0.18$ au. A crucial difference between the case of the Solar system and the one of exoplanetary system is that we know major objects in the Solar system very well but most exoplanets with low masses ($\lesssim M_{\oplus}$) currently remain unknown. In order to include more possible and potential cases, even some of which have not been observed yet, we enlarge the domain $\mathcal{D}$ as well.

\begin{figure}
\centering
\includegraphics[width=.45\textwidth]{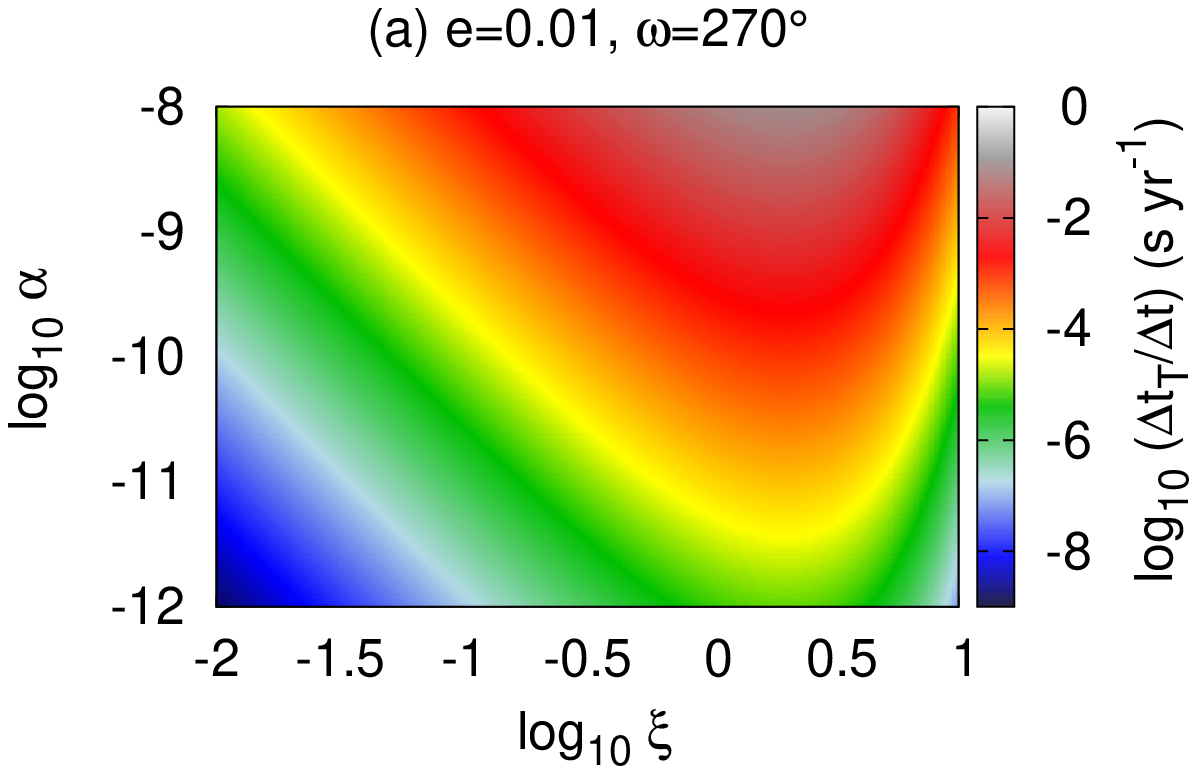}
\includegraphics[width=.45\textwidth]{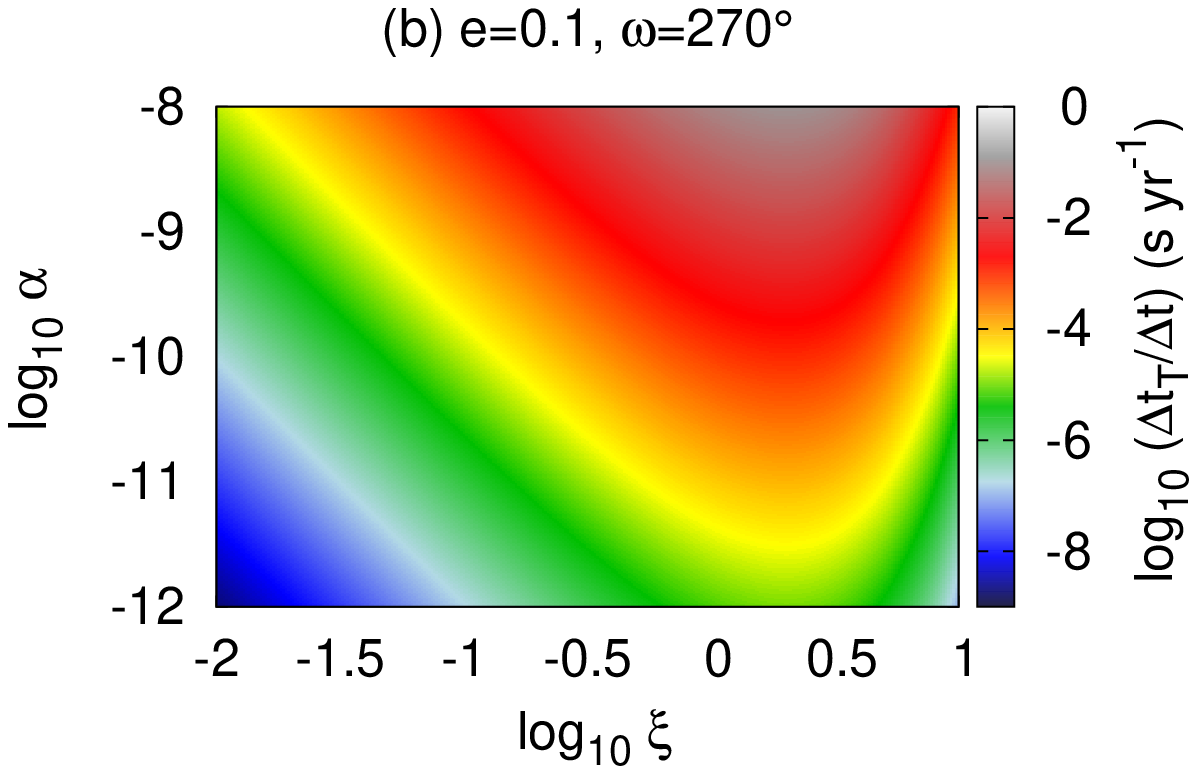}
\includegraphics[width=.45\textwidth]{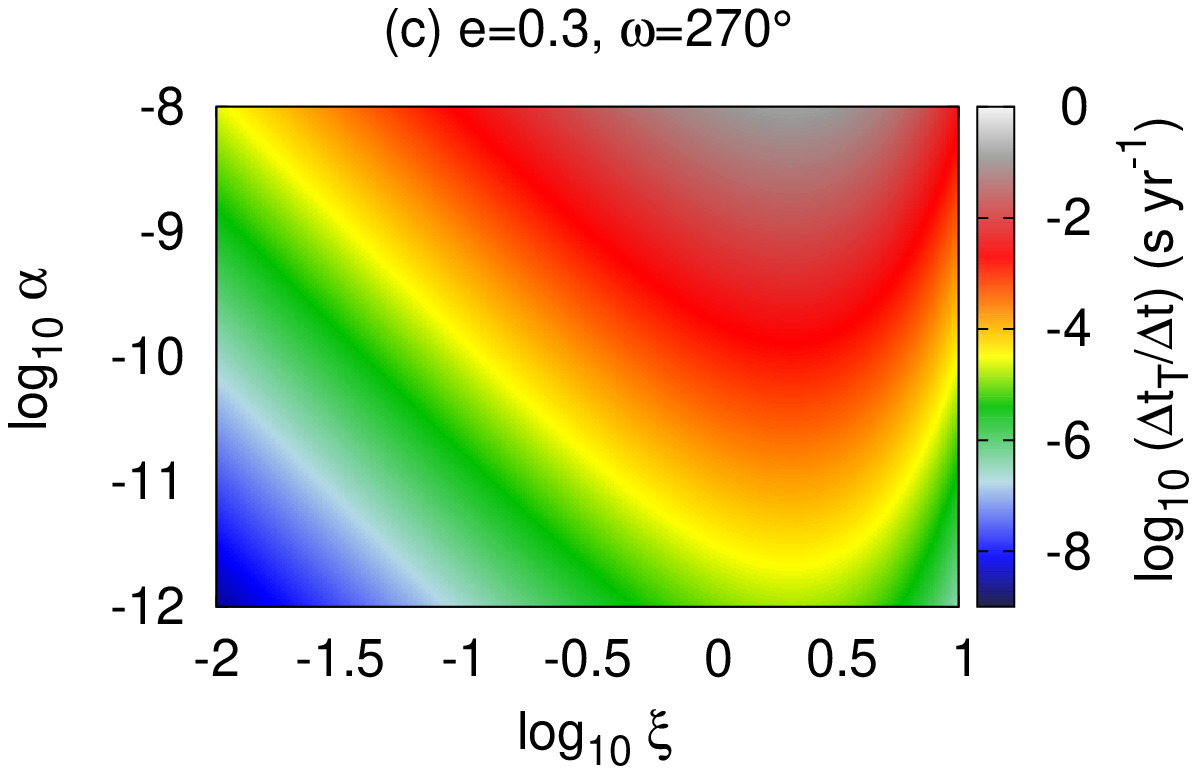}
\includegraphics[width=.45\textwidth]{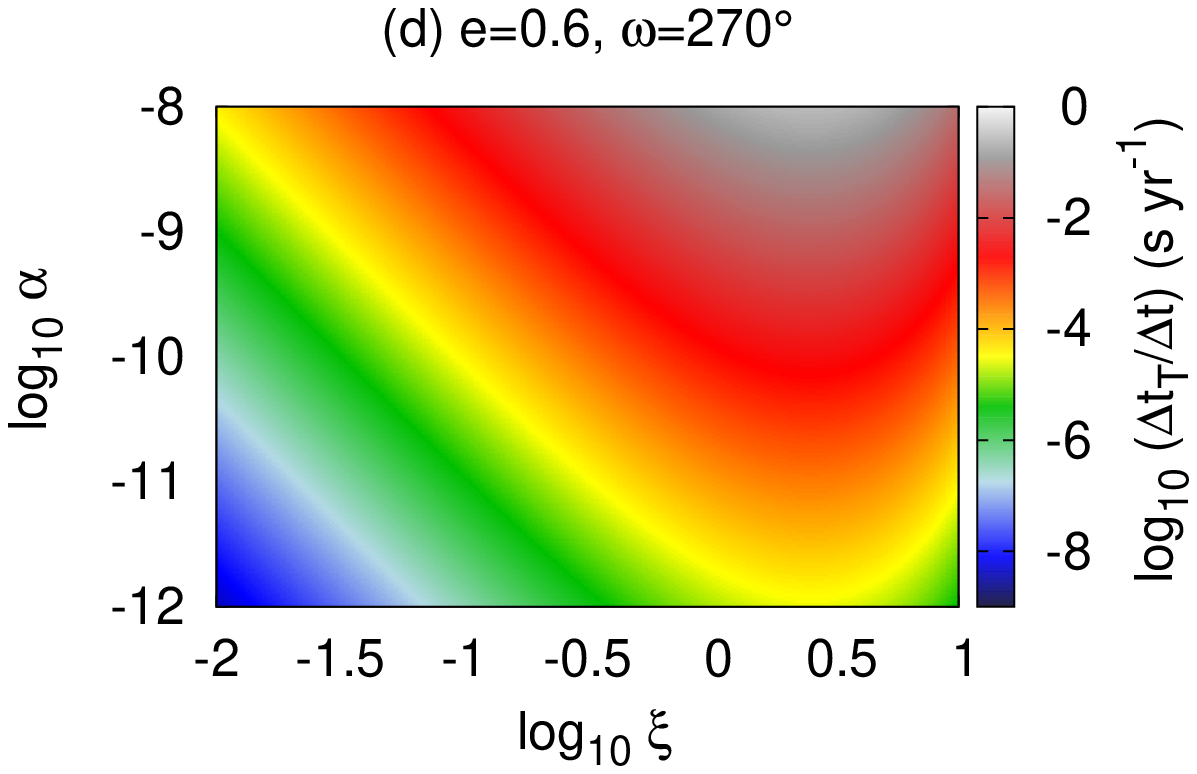}
\caption{\label{fig:1} Color-indexed ${\Delta t_T}/{\Delta t}$ in four cases: (a) $e=0.01$; (b) $e=0.1$; (c) $e=0.3$; and (d) $e=0.6$. These sub-cases share identical logarithmic color bars in the unit of second per year (s yr$^{-1}$) and are all generated by taking $\omega=270^{\circ}$. Their patterns barely change for different values of $\omega$ according to equation (\ref{estvarrho}).}
\end{figure}

Figure \ref{fig:1} shows color-indexed ${\Delta t_T}/{\Delta t}$ in four cases: (a) $e=0.01$; (b) $e=0.1$; (c) $e=0.3$; and (d) $e=0.6$. These sub-cases share identical logarithmic color bars in the unit of second per year (s yr$^{-1}$) and are all generated by taking $\omega=270^{\circ}$ which makes $\varrho_T$ maximum. It can be checked that their patterns barely change for different values of $\omega$ according to equation (\ref{estvarrho}). They tell us that the most optimistic values of Yukawa-type secular TTVs are at the level of $\sim 0.1$ seconds per year. Unfortunately, those values appear only in rarely unique cases (very small regions surrounding $\xi\approx2.2$ and $\alpha=10^{-8}$ in each sub-figures) and, most importantly, they are still at least two orders of magnitude below the current capabilities of observations. Therefore, such deviation from the ISL of gravity is likely too small to detect for the foreseeable future.

There are some other effects that will make the detection more complicated. For example, GR, a stellar quadrupole moment, tidal deformations and an additional planet (perturber) \citep[e.g][]{Miralda-Escude2002ApJ564.1019,Holman2005Sci307.1288,Agol2005MNRAS359.567,Heyl2007MNRAS377.1511,Jordan2008ApJ685.543,Iorio2011MNRAS411.167,Iorio2012CMDA112.117,Nesvorny2012Sci336.1133,Iorio2013CMDA116.357} can also cause secular periastron advances respectively: $\left<\dot{\omega}\right>_{\mathrm{GR}}$, $\left<\dot{\omega}\right>_{\mathrm{quad}}$, $\left<\dot{\omega}\right>_{\mathrm{tide}}$ and $\left<\dot{\omega}\right>_{\mathrm{pert}}$ [see eqs. (1), (3), (5) and (7) in \cite{Jordan2008ApJ685.543} for their expressions]. It is found \citep{Jordan2008ApJ685.543} that $\dot{\omega}_{\mathrm{quad}}$ is usually much less than $\dot{\omega}_{\mathrm{GR}}$; $\dot{\omega}_{\mathrm{tide}}$ and $\dot{\omega}_{\mathrm{pert}}$ are of comparable magnitude to $\dot{\omega}_{\mathrm{GR}}$. The GR periastron advance can result in secular TTVs as \citep{Jordan2008ApJ685.543,Iorio2011MNRAS411.167,Zhao2013RAA13.1231}
\begin{eqnarray}
  \label{GRTTV}
  \frac{\Delta t_T}{\Delta t}\bigg|_{\mathrm{GR}} & = & -3\frac{\sqrt{1-e^2}}{(1+h)^2}\frac{GM_{\ast}}{c^2a} + \mathcal{O}(\cos^2i) \nonumber\\
  & \approx & -\bigg(\frac{47.7\,\mathrm{s}}{1\,\mathrm{yr}}\bigg) \bigg(\frac{M_{\ast}}{M_{\odot}}\bigg)^{2/3} \bigg(\frac{P}{1\,\mathrm{day}}\bigg)^{-2/3} (1 - 2h)\nonumber\\
  & & + \mathcal{O}(\cos^2i,e^2),
\end{eqnarray}
where $c$ is the speed of light and $M_{\ast}$ is the mass of the host star. It means, in a transiting exoplanet system with Sun-like mass, period of $\sim 1$ day and relatively small eccentricity of $\sim 0.1$, the TTVs caused by GR can reach $\sim 40$ seconds in a year. To compare the magnitudes of TTVs triggered by GR and Yukawa correction, we can calculate the ratio as
\begin{eqnarray}
  \label{etaGRYK}
  \eta_{\mathrm{GR/YK}} & \equiv & \frac{(\Delta t_T/\Delta t)_{\mathrm{GR}}}{(\Delta t_T/\Delta t)_{\mathrm{YK}}} = \frac{\left< \dot{\omega} \right> _{\mathrm{GR}}}{\left< \dot{\omega} \right>_{\mathrm{YK}}} \nonumber\\
  & = & 3 \kappa^{1/3} c^{-2} \mu_{\ast}^{2/3} P^{-2/3} f_{1}\, g_{\mathrm{YK}},
\end{eqnarray}
where $\kappa \equiv 4 \pi^2$, $\mu_{\ast}\equiv GM_{\ast}$ and
\begin{eqnarray}
  \label{}
  f_{1} & = & e\,(1-e^2)^{-3/2},\\
  g_{\mathrm{YK}} & = & \frac{\exp(\xi)}{\alpha \xi I_1(\xi e)}.
\end{eqnarray}
Similarly, we also can have three other ratios involving the stellar quadrupole moment, tidal deformations and the perturber as
\begin{eqnarray}
  \label{etaquadYK}
  \eta_{\mathrm{quad/YK}} & \equiv & \frac{(\Delta t_T/\Delta t)_{\mathrm{quad}}}{(\Delta t_T/\Delta t)_{\mathrm{YK}}} \nonumber\\
  & \approx & \frac{3}{2} \kappa^{2/3} J^{\ast}_2 \mu_{\ast}^{-2/3} R_{\ast}^2 P^{-4/3} f_{2}\, g_{\mathrm{YK}} ,\\
  \label{etatideYK}
  \eta_{\mathrm{tide/YK}} & \equiv & \frac{(\Delta t_T/\Delta t)_{\mathrm{tide}}}{(\Delta t_T/\Delta t)_{\mathrm{YK}}} \nonumber\\
  & \approx & 15 \kappa^{5/3}  \mu_{\ast}^{-5/3} P^{-10/3} k_{2,p}\mathcal{T} R_p^5 \frac{M_{\ast}}{M_p} f_3 \, g_{\mathrm{YK}},\\
  \label{etapertYK}
  \eta_{\mathrm{pert/YK}} & \equiv & \frac{(\Delta t_T/\Delta t)_{\mathrm{pert}}}{(\Delta t_T/\Delta t)_{\mathrm{YK}}}\nonumber\\
  & \approx & \frac{3 M_2 P^2}{4 M_{\ast} P_2^2}  f_{2}\, g_{\mathrm{YK}},
\end{eqnarray}
where 
\begin{eqnarray}
  \label{}
  f_{2} & = & e\,(1-e^2)^{-1/2},\\
  f_{3} & = & e(1-e^2)^{-11/2}\bigg[1+\frac{3}{2}e^2+\frac{1}{8}e^4\bigg],\\
  \mathcal{T} & = & 1+\frac{k_{2,s}}{k_{2,p}}\bigg(\frac{R_{\ast}}{R_p}\bigg)^5\bigg(\frac{M_p}{M_{\ast}}\bigg)^2.
\end{eqnarray}
Here, $J^{\ast}_2$ is the stellar quadrupole moment, $R_{\ast}$ is the radius of the host star; $M_p$ is the mass of the planet, $R_p$ is the radius of the planet; $k_{2,s}$ and $k_{2,p}$ are the apsidal motion constants for the star and planet respectively, which depend on the mass concentration of the tidally deformed bodies \citep{Sterne1939MNRAS99.451}; $M_2$ is the mass of the second planet and $P_2$ is the Keplerian period of its orbit. In the calculations of these ratios [equations (\ref{etaGRYK}) and (\ref{etaquadYK})--(\ref{etapertYK})], equation (\ref{Yukawadomegadt}) in this work and eqs. (1), (3), (5) and (7) in \cite{Jordan2008ApJ685.543} are used. 

\begin{figure}
\centering
\includegraphics[width=.43\textwidth]{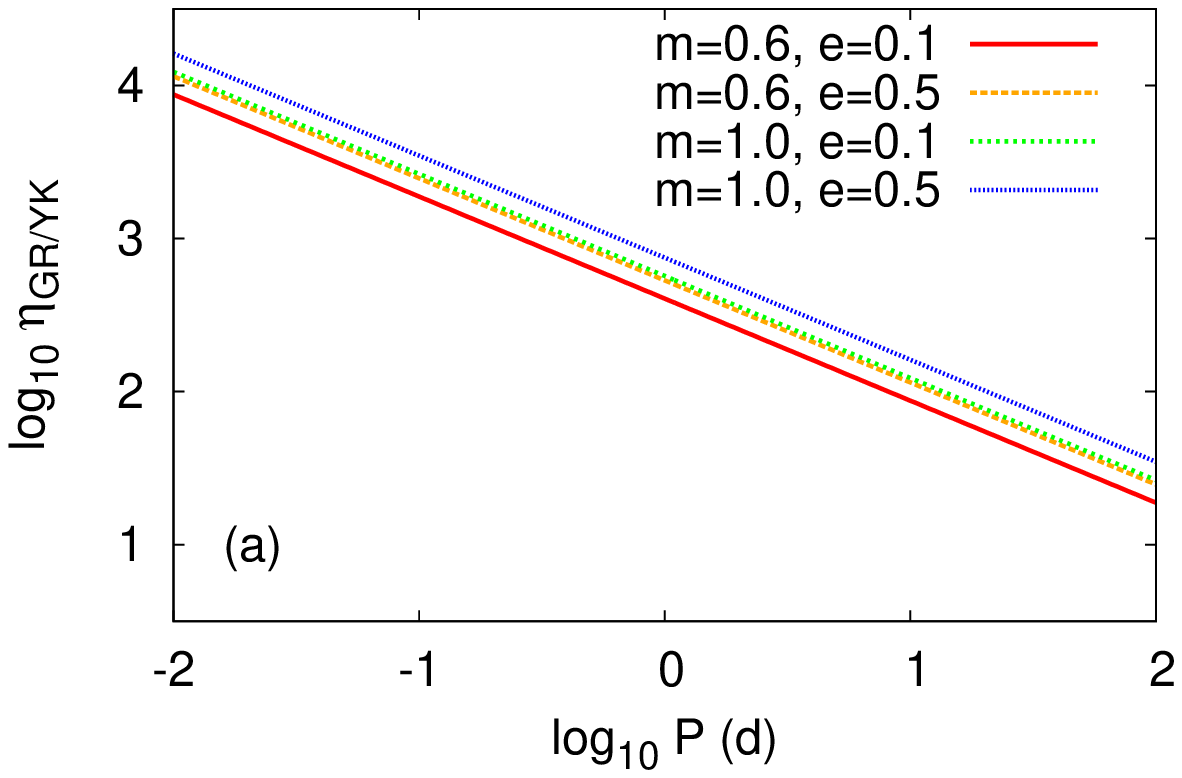}
\includegraphics[width=.43\textwidth]{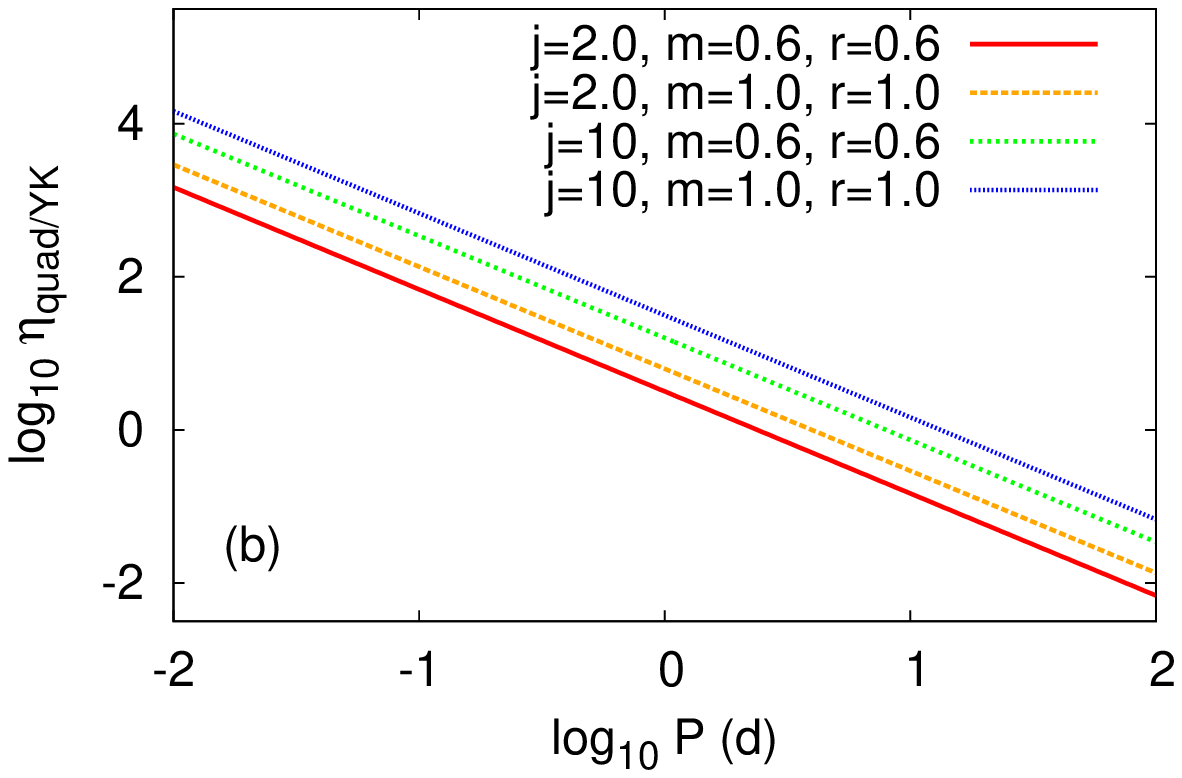}
\includegraphics[width=.43\textwidth]{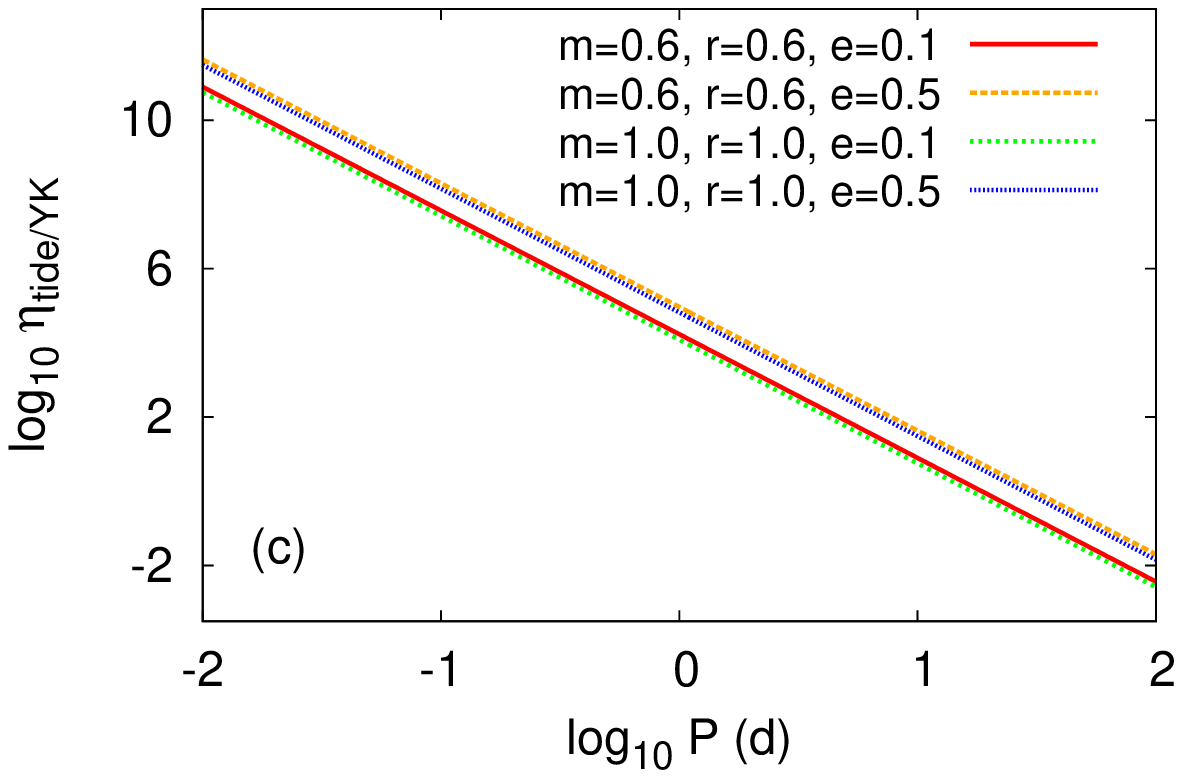}
\includegraphics[width=.43\textwidth]{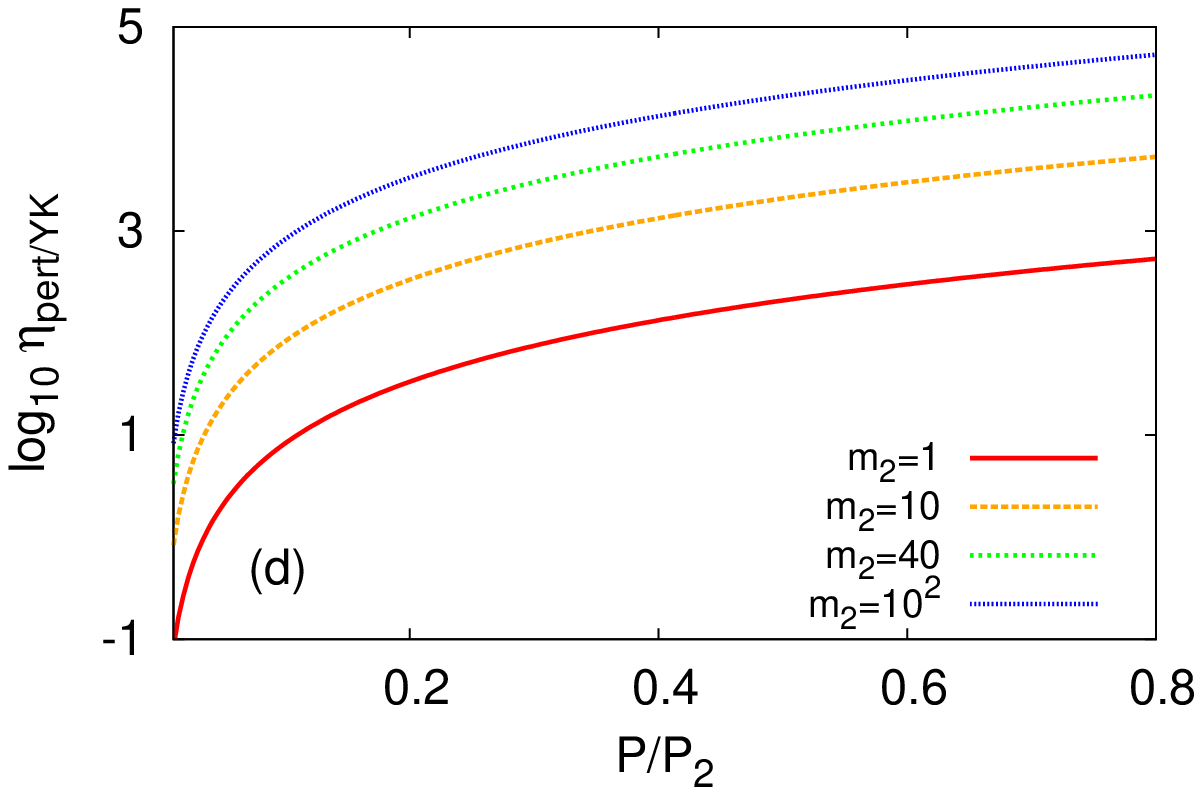}
\caption{\label{fig:2} The curves of $\eta_{\mathrm{GR/YK}}$, $\eta_{\mathrm{quad/YK}}$, $\eta_{\mathrm{tide/YK}}$ and $\eta_{\mathrm{pert/YK}}$ are shown in the panels of (a), (b), (c) and (d) respectively. We fix $\xi=2.2$ and $\alpha=10^{-8}$ for all of them. In panel (a), $m\equiv M_{\ast}/M_{\odot}$; in (b), we take $e=0.3$ and have definitions as $j=J^{\ast}_2\times 10^{7}$ and $r\equiv R_{\ast}/R_{\odot}$; in (c), we assume a Jupiter-like giant planet with $k_{2,p}=0.25$ and a host star with $k_{2,s}=0.01$; and in (d), we define the mass of a perturber as $m_2\equiv M_{2}/M_{\oplus}$.}
\end{figure}

In order to estimate these ratios, we fix $\xi=2.2$ and $\alpha=10^{-8}$ because they generate the most optimistic values of Yukawa-type secular TTVs, which are at the level of $\sim 0.1$ seconds per year (see figure \ref{fig:1}). After evaluating them, we find secular TTVs caused by GR, the stellar quadrupole moment, tidal deformations and a perturber are usually larger than the one due to the Yukawa correction by several orders of magnitude (see figure \ref{fig:2}). Figure \ref{fig:2}(a) shows $\eta_{\mathrm{GR/YK}}$ with respect to $P$, where $m\equiv M_{\ast}/M_{\odot}$. It indicates, for an exoplanet with $P\sim 10$ days, the TTVs caused by GR are about $10^2$ times greater than those triggered by possible deviation from ISL of gravity. The comparison of TTVs by the stellar quadrupole moment and Yukawa correction is given in figure \ref{fig:2}(b) in which $j\equiv J^{\ast}_2\times 10^{7}$, $r\equiv R_{\ast}/R_{\odot}$ and $e$ is fixed as $0.3$. It suggests effects from the quadrupole moment will suppress the Yukawa correction on a close-in planet. Some curves of $\eta_{\mathrm{tide/YK}}$ are presented in figure \ref{fig:2}(c) where we assume a Jupiter-like giant planet with $k_{2,p}=0.25$ by a polytrope of index $n\approx 1$ \citep{Hubbard1984BookPI} and a host star with $k_{2,s}=0.01$ \citep{Claret1992A&AS96.255}. Like the trends in figure \ref{fig:2}(b), the TTVs by tidal deformations are much larger than those by Yukawa correction for hot-Jupiters. In figure \ref{fig:2}(d), these curves show the relative strength of TTVs by a perturber and those by Yukawa correction, where $m_2\equiv M_{2}/M_{\oplus}$. The influence of a perturber can reach the level of several orders of magnitude greater than the effects of Yukawa correction. For a perturber with $\sim M_{\oplus}$, its contribution in TTVs can be about 10 times larger than Yukawa-correction's when $P:P_2\approx 1:2$. However, such perturbers with low masses are currently difficult to detect and remain unknown in most cases. The resulting uncertainties will ruin any efforts to test ISL of gravity using exoplanetary TTVs and they will likely be exceptionally difficult to remove from a signal that should be seen. Separating, discriminating and extracting various contributions in TTVs for future positive detection of possible deviation from ISL require tremendous advances of techniques for observations and sophisticated methods of data analysis.

\section{Conclusions and discussion}

\label{sec:con}

In the context of potential and considerable increase of the number of transiting exoplanets and the precision of measured times of transit minima by ground-based and space-borne observatories used for studying exoplanet transits  now and in the future, we study the possibility of testing fundamental laws of nature in these system via TTVs. Focusing on presumable violations of the ISL of gravity which are parameterized by the fifth-force-like Yukawa-type correction to the Newtonian gravitational force, we investigate their effects on secular TTVs and analyze their observability. It is found that the most optimistic values of Yukawa-type secular TTVs are at the level of $\sim 0.1$ seconds per year. Those values unfortunately appear only in rarely unique cases and, most importantly, they are still at least two orders of magnitude below the current capabilities of observations. Such deviation from the ISL of gravity is likely too small to detect for the foreseeable future.

Moreover, exoplanetary systems are full of complexity so that many sources can trigger secular TTVs, such as GR, a stellar quadrupole moment, tidal deformations and a perturber. After calculating the ratios between TTVs by Yukawa correction and those caused by these four effects, we find the signals of Yukawa correction are much weaker than others. The uncertainties of perturbers with low masses, which are usually unknown for now, will ruin any efforts to test ISL of gravity using exoplanetary TTVs and they will likely be exceptionally difficult to remove from a signal that should be seen. Separating, discriminating and extracting various contributions in TTVs for positive detection require tremendous advances of techniques for observations and sophisticated methods of data analysis.

\section*{Acknowledgments}

We acknowledge very useful and helpful comments and suggestions from our anonymous referee. The work of YX is supported by the National Natural Science Foundation of China Grant No. 11103010, the Fundamental Research Program of Jiangsu Province of China Grant No. BK2011553 and the Research Fund for the Doctoral Program of Higher Education of China Grant No. 20110091120003. The work of XMD is funded by the Natural Science Foundation of China under Grant No. 11103085 and  the Fundamental Research Program of Jiangsu Province of China Grant No. BK20131461.

\bibliographystyle{mn2e.bst}
\bibliography{Gravity20131129.bib}

\end{document}